\documentstyle[twocolumn,aps]{revtex}
\tightenlines
\begin{document}

\newcommand{\beq}{\begin{equation}}
\newcommand{\eeq}{\end{equation}}
\newcommand{\bea}{\begin{eqnarray}}
\newcommand{\eea}{\end{eqnarray}}
\newcommand{\ba}{\begin{array}}
\newcommand{\ea}{\end{array}}
\newcommand{\om}{(\omega )}
\newcommand{\bef}{\begin{figure}}
\newcommand{\eef}{\end{figure}}
\newcommand{\no}{\nonumber}
\newcommand{\etal}{{\em et~al }}
\newcommand{\cf}{{\it cf.\/}\ }
\newcommand{\ie}{{\it i.e.\/}\ }
\newcommand{\eg}{{\it e.g.\/}\ }

\title{Continuous variable quantum cryptography \\ using coherent states}
\author{Fr\'ed\'eric Grosshans and Philippe Grangier}
\address{Laboratoire Charles Fabry de l'Institut d'Optique 
(CNRS UMR 8501)
F-91403 Orsay, France}

\maketitle

\begin{abstract}

We propose several methods for quantum key distribution (QKD)  
based upon the generation and transmission of random distributions of coherent 
or squeezed states, and
we show that they are are secure against individual eavesdropping attacks.
These protocols require that the transmission of the optical line between Alice and Bob
is larger than  50 $\%$, but they do not rely on ``non-classical" features such as squeezing.
Their security is a direct consequence of the no-cloning theorem, 
that limits the signal to noise ratio of possible quantum measurements
on the transmission line. Our approach can also be
used for evaluating various QKD protocols using light with gaussian statistics.

PACS numbers:  03.67.Dd, 42.50.Dv, 89.70.+c 
\vspace{0.25cm}

\end{abstract}

Since the experimental demonstration of quantum teleportation
of coherent states \cite{Fu}, a lot of interest has arisen in continuous
variable quantum information processing. 
In particular, a stimulating question is whether
quantum continuous variables (QCV) may provide 
a valid alternative to the usual ``single photon"
quantum key distribution schemes \cite{QKD}.
Most present proposals to use QCV for QKD 
\cite{GP01,Hillery00,CLvA00,CIvA01,vACC01,Ralph00a,Ralph00b,Reid00,SKL01,NGL01,LSKWL01,BSJL01,POK00}.
are based upon the use
of ``non-classical" light beams, such as squeezed light, or 
pairs of light beams that are
correlated for two different quadratures components 
(the so-called ``EPR" beams, by analogy
with the historical paper by Einstein, Podolski and Rosen \cite{epr}).
But recent work on this subject \cite{GG01} underlined the crucial importance
of the continuous variable version of the 
no-cloning theorem \cite{cerf}, as soon as security is 
concerned in any exchange using QCV. 

In this letter, we show that there is actually no need for 
squeezed light : an equivalent level of security may be obtained
by simply generating and transmitting random distributions of coherent states.
The security of this novel protocols is related to the no-cloning theorem,
that limits possible eavesdropping even though the transmitted light
has no ``non-classical" feature such as squeezing.
We show that our analysis can be also applied to other protocols using 
light with gaussian statistics, \ie squeezed or EPR beams,
making thus the comparison easier.
The basic tools for this analysis are the ones 
that have been extensively used for linearized quantum optics,
including in particular optical quantum non-demolition 
(QND) measurements \cite{qnd}.
Before presenting our protocol, we will briefly review the current 
literature on continuous variables QKD.

Here we consider security against individual attacks
only, and we do not address the issue of unconditionnal security, that was
demonstrated by Gottesman and Preskill \cite{GP01} for squeezed states protocols 
(unconditional security of coherent states protocols remains an open question).
Security against individual attacks was previously considered by many authors.
Hillery proposed a QKD scheme  based on binary modulated squeezed light \cite{Hillery00}.
Cerf \etal showed it could be improved considering gaussian modulation \cite{CLvA00,CIvA01}
and described a reconciliation protocol \cite{CIvA01,vACC01} to implement this
improved protocol. In the present work we will
generalize this approach to the various single beam protocols of the litterature
\cite{Hillery00,CLvA00,CIvA01,vACC01,Ralph00a,Ralph00b,Reid00,SKL01,NGL01}.
The protocol described in \cite{CLvA00,CIvA01} is then a particular member of the
family of protocols described here.
EPR beams were also considered for QKD schemes. 
Some schemes need the propagation of one beam only from Alice to Bob 
\cite{Ralph00b,Reid00,SKL01,NGL01,LSKWL01,BSJL01},
the other half of the EPR pair being measured by Alice,
whereas others need the propagation of two modes (or more) of the electromagnetical
field  \cite{Ralph00a,LSKWL01,BSJL01,POK00}.
In the first family, Reid \cite{Reid00} and Ralph \cite{Ralph00b} consider 
``binary" modulated EPR beams, created by a parametric amplifier with a modulated
seed \cite{Reid00} 
or interfering modulated squeezed beams\cite{Ralph00b}, whereas 
Silberhorn \etal \cite{SKL01,LSKWL01} and Navez \etal \cite{NGL01} extract their key
from correlated measurement sequences. 
As we will show below, 
these schemes can be viewed as the transmission of a modulated sub-shotnoise beam.
Bencheikh \etal \cite{BSJL01} extract the binary key
directly from the gaussian correlations. 
This extraction can be optimized using the reconciliation protocol described in
\cite{CIvA01,vACC01}.
The protocols transmitting several quantum-correlated modes of the electromagnetic
field, using two beams \cite{Ralph00a,LSKWL01,BSJL01,POK00}
are beyond the scope of this letter, because their security analysis should take
into account simultaneous attack on both modes. 
However, similar gaussian extension of these protocols seem possible.
Finally, Ralph examined a binary modulated coherent beam protocol
\cite{Ralph00a,Ralph00b},
and showed its need for privacy amplification \cite{Ralph01}. 
Here we will introduce a family of gaussian protocols, and we will
show that the coherent state version is secure and as
efficient as the corresponding squeezed light or EPR protocols.

{\it General principle of the protocols.}
The QKD protocols we study here are single gaussian beam protocols.
Alice modulate randomly a gaussian beam and send it to Bob through a gaussian noisy
channel. 
Both phase and amplitude are modulated with gaussian random numbers, 
since it allows an optimal information rate \cite{Sha48}. 
Bob then measures either the phase or the amplitude of this beam and informs Alice
which measurement he made.
Bob and Alice have then two correlated sets of gaussian variables, from which they
can extract a common secret string of bits as explained below. 

The basic tool that we will use is the Shannon formula giving the optimum
information rate $I$ of a noisy transmission channel,
in units of bits/symbol \cite{Sha48}. If the noise is white and gaussian 
and the signal to noise ratio (SNR) is $\Sigma$, this optimum information rate  is
\beq
I_{AB}=1/2 \log_2 (1 + \Sigma).
\label{sh}
\eeq
Since this optimum can be closely approached only if the signal has a gaussian statistics
\cite{Sha48}, we will consider only gaussian modulation protocols, and use
(\ref{sh}) to calculate the amount of private information that
Alice and Bob may exchange in presence of the eavesdropper Eve.

The sliced  reconciliation protocol described in detail in
\cite{CIvA01,vACC01} and briefly sketched in the Appendix
allows us to get arbitrarly close to the value given by (\ref{sh}). 
For security purposes, one must assume that
Eve has an arbitrary powerful computer, and thus she is able to reach this limit. 
In case Alice and Bob are not, 
they will have to allow for an extra security margin (see {\em Discussion} below).
We note that it is not required to specify a ``digitizing step" to connect
the continuous variable and a bit value: 
the bits will appear at the end of the reconciliation protocol \cite{CIvA01,vACC01}.
At this stage, Alice and Bob share  a string of bits which is partly known by Eve. 
They can then use standard privacy amplification protocol \cite{PrivacyAmp} to agree
on a secret key. The rate at which this secret key can be constructed is
\begin{equation}
         \Delta I=I_{AB}-I_{AE},
         \label{DeltaI}
\end{equation}
where $I_{AB}$ ($I_{AE}$) is the information rate between Alice and Bob (Eve).

{\it Eavesdropping.}  The $I_{AB}$ term of (\ref{DeltaI}) is easy to compute for a
given scheme, 
the signal to noise ratio $\Sigma_B$ being known. We have to assume $I_{AE}$ being
the 
maximum possible given the laws of physics
(considering only individual attacks, coherent attacks are beyond the scope of this
letter). 
If the protocols are globally invariant under the
exchange of the two quadratures $X$ and $P$, the best tactic for Eve is to keep
this property in her attacks.
Therefore, we can restrict us to attacks that treat equally both
quadrature without loss of generality.

Given these hypothesis, we will use a general result, that is demonstrated in
\cite{GG01} :
if the added noise on Bob's side is  $\chi N_0$, where $N_0$ is the vacuum noise
variance, then the minimum added noise on Eve's side is $\chi^{-1} N_0$.
This applies to both quadratures, and the 
added noise may be due to line losses, eavesdropping, or any other reason
\cite{GG01}.
Since the demonstration of ref. \cite{GG01} is just another form of the 
no-cloning theorem, it also adresses any individual attack by Eve
using a cloning machine \cite{cerf}. 
When the line has a transmission $\eta$ with no Eve present, 
one has $\chi=(1-\eta)/\eta$. The best attack for Eve is then to take
a fraction $1-\eta$ of the beam at Alice's site, and to send the fraction $\eta$ to Bob through
her own lossless line (that may be a perfect teleporter). Eve is then totally undetected,
and she gets the maximum possible information according to the no-cloning theorem. 

Equation (\ref{DeltaI}) shows that these protocols are secure as long as Bob has 
a more in formation on Alice's key element than Eve, 
\ie as long as $I_{AB}>I_{AE}$.
Since the Shannon formula (\ref{sh}) is valid for both Bob and Eve, 
the security condition is just a condition on the signal to noise ratios, 
which turns to be a condition on the added noises, 
since the signal and the noise added at Alice's side (quantum noise, Alice's
technical noises) are the same.
\beq
  \Delta I >0 \; \Leftrightarrow \;  \Sigma_B > \Sigma_E \;  \Leftrightarrow \;  \chi <1           
\eeq
Since  $\chi=(1-\eta)/\eta$ for a line with transmission $\eta$, 
the condition $\chi <1$ requires that $\eta > 1/2$. 
Therefore, a usable key can be obtained in principle as soon as the 
transmission losses are less 3dB. Taking into account the standard loss of 0.2dB/km
in optical fibers at 1550 nm, the typical range would be around 10~km.

In this security evaluation, the noise added in Alice's side cancels out 
because it disturbs equally Eve and Bob. 
This `cancelled' noise includes the quantum noise of the beam.
As a consequence, the security of these protocols relies of the quantum aspects of 
measuring or copying,
but not on any quantum feature of the beam, like squeezing or entanglement. 
We can do quantum cryptography with coherent beams, as mentionned by Ralph
\cite{Ralph00a,Ralph00b} or even with highly noisy beams. 
Quantum features of the beams might influence some characteristics of the protocol 
like the secret key rate 
or the amount of classical communication needed to agree on the secret key, 
but not its security.

{\it Coherent Beam protocol.} 
Let us now explicitly describe the coherent beam protocols of this family: 

1. Alice draws two random numbers $x_A$ and $p_A$ from a gaussian law with variance
$V_A N_0$

2. She sends to Bob the coherent state $\left|x_A + i p_A\right>$

3. Bob randomly chooses to measure either $X$ or $P$. This measurement can be done
perfectly.

4. Using a classical public channel he informs  Alice  about the observable that he
measured
(like in the BB84 protocol, half of the key generated by Alice is unused)

5. Alice and Bob share two correlated gaussian variables. Then they may use
the ``sliced reconciliation" protocol \cite{vACC01,CIvA01} to transform it into
errorless bit strings. Finally, they have to use a standard protocol for
privacy amplification \cite{PrivacyAmp} in order to distill the private key. 

According to eq. (\ref{sh}), the channel rate $\Delta I$ for the private key will
be: 
\begin{equation}
   \textstyle
   \Delta I =\frac12\log_2 (1+\Sigma_B) - \frac12\log_2(1+\Sigma_E) 
\end{equation}
The total variance of any quadrature of the beam when it leaves Alice's realm is
$VN_0=V_AN_0+N_0$.
Using the expressions $1+\Sigma_B = \frac{V+\chi}{1+\chi}$, and 
$1+\Sigma_E = \frac{V+1/\chi}{1+1/\chi}$,
the useful secret information rate is :
\begin{eqnarray}
  \label{DeltaICoh}
   \textstyle
   \Delta I =\frac12\log_2\frac{V+\chi}{1+V\chi}
\end{eqnarray}
If $\chi<1$, $\Delta I$ will increase as a function of the signal modulation $V_A$.
For large modulation ($\chi V \gg 1$), the asymptotic value of $\Delta I$ is :
\beq
  \textstyle
  \Delta I_{asymp} =  -\frac{1}{2} \log_2 \chi = \frac{1}{2}
             \log_2\frac{\eta}{1-\eta}
  \label{DeltaIcohAsymp}
\eeq
while the raw channel rate between Alice and Bob is  $I_{AB} = \frac{1}{2} \log_2
(V/(1+\chi))$.

{\it Squeezed state protocol.}
This protocol can straightforwardly be generalized to
modulated squeezed beam, with a squeezing factor $s<1$.
The protocol becomes :

1. Alice chooses randomly if the beam is squeezed in $X$ or $P$
(for instance we will later assume the beam being $X$-squeezed). 
Let denote $\left|\psi\right>$ this squeezed state.

2. Alice draws two random numbers $x_A$ and $p_A$ from two gaussian laws
with variances
 $V_{x_A}N_0$ and $V_{p_A}N_0$. 
The two squeezed direction are indistinguishable for Eve iff
\begin{equation}
  \textstyle
  V_{x_A}N_0+sN_0= V_{p_A}N_0+\frac1sN_0\equiv VN_0
\end{equation}

3. Alice sends to Bob the displaced squeezed state $D(x_A+ip_a)\left|\psi\right>$

4. Bob randomly chooses to measure either $X$ or $P$.

5. Using a public channel, Alice and Bob inform each other about the squeezing direction
and the measured observable.

6. Like with coherent states Alice and Bob share correlated gaussian
variables, from which they can extract a private binary key.

This protocol obviously reduces to the protocol described above if $s=1$. 
Another limit, where $V_{p_A}=0$ or $V=1/s$, is the protocol
described by Cerf \etal in \cite{CLvA00,CIvA01}. In this case, information is
gathered for the key only when Bob makes the right guess.

To compute the private rate $\Delta I$, we will
average between the right guesses and the wrong guesses :
\begin{eqnarray}
   \textstyle
   \Delta I &=& \textstyle \frac12[(I_{ABX}-I_{AEX})+(I_{ABP}-I_{AEP})]\\
            &=&  \textstyle \frac14\log_2\frac{(1+\Sigma_{BX})(1+\Sigma_{BP})}{(1+\Sigma_{EX})(1+\Sigma_{EP})}
\end{eqnarray} 
We have $ \Sigma_{BX} = \frac{V_{x_A}}{s+\chi}=\frac{V-s}{s+\chi}$ and 
$1+\Sigma_{BX} = \frac{V+\chi}{s+\chi}$.
The three other signal to noise ratios are obtained by replacing $\chi$ or/and $s$
by $\chi^{-1}$ or $s^{-1}$. Therefore,
\begin{eqnarray}
   I_{AB}&=&\textstyle \frac14\log_2\frac{(V+\chi)^2}\chi
                      - \frac14\log_2(\chi+\frac{1}{\chi}+s+\frac{1}{s} )\\
  I_{AE}&=&\textstyle  \frac14\log_2\frac{(V+1/\chi)^2}{1/\chi}
                     - \frac14\log_2(\chi+\frac{1}{\chi}+s+\frac{1}{s})
\end{eqnarray}
Since the $s$-dependent term of these information rates are the same, they cancel
each other in $\Delta I$. 
The secret information rate is thus again given by eq.  (\ref{DeltaICoh}),
and does not depend on the degree of squeezing.

{\it Extension to EPR case.}
The previous description does not apply directly on EPR protocols.
However, an EPR QKD protocol where Alice keeps one of the beams and sends the other
to
Bob is logically equivalent to a randomly modulated beam with a sub-shot noise
quantum variance.
Let note $X_A$ the quadrature Alice measures and $X_{out}$ the same quadrature of
the beam sent to Bob
when it leaves Alice's lab. For a standard non-modulated 
EPR scheme \cite{SKL01} we have the following relations :
\begin{eqnarray}
  \left<X_A^2\right>=\left<X_{out}^2\right>&\equiv&V = (s+1/s)/2
     \label{EPRVar}\\
  \left<(X_A-X_{out})^2\right>&=&2s\\
  \left<X_AX_{out}\right>&=&V-s  
     \label{EPRCorr}
\end{eqnarray}
We can separate Bob's beams in two parts, that are respectively correlated
and uncorrelated with Alice's measurement, by writing
 $X_{out} = gX_A+N$ where $\left<X_AN\right> = 0$.
Bob's beam is then equivalent to a beam with quantum noise
$\left<N^2\right>$ on quadrature $X$, 
which is randomly modulated with the variable $gX_A$. 
Using eqs (\ref{EPRVar},\ref{EPRCorr}) one gets:
\begin{eqnarray}
 & g& = 1- s/V = (1-s^2)/(1+s^2) \\
               & \left<N^2\right> &= s (2-s/V)= 2s/(1+s^2) .
\end{eqnarray}
These equations describe the case where Alice and Bob measure the same quadrature.
When Alice changes her quadrature, while Bob keeps the same measurement,
the initial wave packet is reduced onto a noisy quadrature, 
and no useful correlation is generated. On the average, 
the information rate is therefore half of the ``equivalent" modulation scheme.
Using (\ref{EPRVar}), we have then: 
\begin{eqnarray}
  1+\Sigma_B&=&
             \textstyle  1+\frac{g^2 V}{\left<N^2\right>+\chi}=\frac{V(V+\chi)}{1+\chi V}\\
  \Delta I&=&
    \textstyle \frac14\log_2(\frac{V+\chi}{1+\chi V}\frac{1+V/\chi}{V+1/\chi}) =
\frac12\log_2(\frac{V+\chi}{1+\chi V})
\label{DIepr}
\end{eqnarray}
This value of $\Delta I$ is again just the same as
the coherent state result (\ref{DeltaICoh}) for given $\chi$ and $V$,
so that $s$ is defined by (\ref{EPRVar}).
Adding excess noise or a modulation on the outgoing beam
brings no further improvement.

{\it Discussion.} 
Various comments are in order. First, it appears that 
non classical features like squeezing or EPR correlations have no
influence on the achievable secret key rate for the family of protocols that were
described here. This result may not apply to all possible protocols,
\eg, we did not consider using a continuous quantum memory.
On the other hand, since the raw information rate are different for the same secret key rate, 
squeezed beams can be used to save classical communications during the privacy
amplification procedure. The EPR beams have also the advantage of directly providing
quantum-generated gaussian noise, rather than having it externally generated by Alice. 
More importantly,  entanglement, that is not directly used in the present
protocols, can be useful to beat the 3 dB limit by using more than one beam.
Though the 3 dB loss limit of our cryptography protocols makes 
their security demonstration quite intuitive, there exist multiples ways 
for Alice and Bob to go beyond this limit.
The  most radical way is to send many EPR beams through the noisy channel, 
then to use entanglement purification \cite{Dal00}
to build stored entanglement between Alice and Bob,
and finally to implement a high fidelity teleporter. 
For any finite value of the losses and EPR entanglement, an arbitrarily high fidelity
can be achieved \cite{Dal00}.
The no-cloning theorem ensures the security of these schemes as soon as the fidelity
of the teleporter is above 2/3 \cite{GG01}, which is equivalent to the 3~dB loss
limit discussed above. In some sense, a ``lossless" line is re-created by using
entanglement purification.   
There may exist more realistic ways to cross the 3~dB barrier. 
For instance, Alice and Bob may ``invert" the reconciliation procedure,
with Alice guessing Bob's measurement instead
of Bob guessing Alice's value \cite{PrivacyAmp}. 
This inverted procedure may be more efficient, but 
its complete security analysis is beyond the scope of this letter.

On the practical side, one should note that Bob's detectors are not ideal, but have 
a non-zero electronic noise $B_0$, that should be much smaller than $N_0$,
and a maximum (saturation) input power $\sigma B_0\gg N_0$, where $\sigma \gg1$ is
the detector's dynamics. Taking into account these characteristics
in the simplest coherent state protocol
gives an optimum value of the signal variance, $V_A \sim \sqrt{\sigma}$. 
Another point is that Alice and Bob may not be able to achieve
the Shannon limit (\ref{sh}), due to limited computing power (no such limitation
is relevant for Eve). Assuming that the effective information rate between
Alice and Bob is reduced by a factor $\alpha < 1$, the net secret rate 
becomes $\Delta I_{eff} = \alpha I_{AB} - I_{AE}$, 
and  remains positive if  $\alpha > I_{AE}/I_{AB}$.
The quantity $\Delta I_{eff}$ is plotted on Fig.1 for $\alpha = 1$ (full lines), 
and for various values of $\alpha$ that are arbitrarily associated with
various values of the SNR (dashed lines). It is clear from that figure that low values of 
$\alpha$ reduce the transmission range in which the protocol is secure.
We note that according to \cite{CIvA01,vACC01}, the sliced reconciliation protocol
should yield $\alpha \sim 1$ (see also Appendix), but this may be costly in terms of 
calculation time and public channel transmissions.
All these constraints should eventually be taken into account to choose
the most appropriate value of $V_A$.

As a conclusion, it is possible to design a QKD scheme with coherent states, secure
against any individual attack, by using optimized reconciliation protocols and 
privacy amplification. 
Since the protocol does not require squeezing, it can
be implemented by sending light pulses in a low-loss optical fiber, like
in a coherent optical telecommunication scheme. 
In that case, all pulses will be useful,
but half of the random numbers generated by Alice will not be used.
We  demonstrated that the protocol is 
asymptotically secure \cite{vACC01} for losses smaller than 3dB
(or a teleportation fidelity larger than 2/3 \cite{GG01}), and the net information
rate for the private key with a large signal modulation is 
$1/2 \log_2 (1/\chi) = 1/2 \log_2 (\eta/(1-\eta))$.

\subsection*{Appendix : Sliced reconciliation protocol}

In the n-slice version of the reconciliation protocol proposed in ref. \cite{vACC01},
the real axis representing the amplitude of the signal is split in $2^n$ intervals
$s_1 = ]- \infty , \; -t_1], \; s_2 = ] -t_1, \; -t_2], \; ... \; s_{2^n} = ]t_{2^n-1}, \;
+\infty [$, where $t_p = - t_{2^n -p}$, and  $t_{2^{n -1}} = 0$.
Alice assigns an amount of $n$ bits to an amplitude that lies in the interval $s_p$,
by using the parity of $p$ for bit 1,
of $Floor(p/2)$ for bit 2, ... , and  of $Floor(p/2^{n-1} )$ for bit $n$.
After receiving the data, 
Bob makes an optimized guess of the first bit value using appropriate weighting
functions, that are computed by optimizing the choice of the  $\{ t_p \}$
(this optimization is made only once, before exchanging the data).
After a first correction round by exchanging public data between Alice and Bob, 
Bob knows the correct value of the first bit.
Then he tries to guess the second bit, with a much higher probability of success,
because he already knows the first one. 
By increasing both the SNR $\Sigma$ and the number of slices, the process gets more and more 
efficient, keeping the same main idea :
after each correction round, Bob can guess the next bit with
a higher probability. 
For the 5-slice protocol with $\Sigma = 15$ presented in
\cite{vACC01}, the probabilities of guessing right for slices 4 and 5 are
respectively 0.976
and 0.999994, and the efficiency is more than 90\% of the Shannon limit 
$\frac12\log_2(16) = 2$. 

{\it Acknowledgments.} 
This work was carried out in the framework of the European
IST/FET/QIPC project ``QuICoV".
We are grateful to N.J.~Cerf and G. Van Assche for helpful discussions.


\begin{figure}
\vspace{6cm}
\vspace{0.5cm}
\caption{Private channel information rate $\Delta I$ as a function of the channel noise $\chi$.
The three curves in full lines correspond to $V_A = 1, 5, 50$ from the bottom to the top,
assuming that the reconciliation protocol between Alice and Bob reaches the Shannon limit. 
The three curves in dashed lines correspond to the effective $\Delta I$ with the sames values of $V_A$,
with (arbitrarily chosen) reconciliation efficiencies $\alpha$ that are respectively 
0.6, 0.8 and 0.95 of the Shannon limit.}
\label{fig1}
\end{figure}

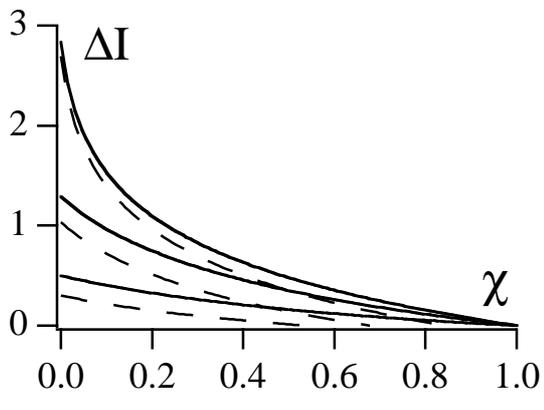
\end{document}